\documentclass[twocolumn,useamsfonts,pdftex]{pasj01}
\Received{$\langle$reception date$\rangle$}
\Accepted{$\langle$acception date$\rangle$}
\Published{$\langle$publication date$\rangle$}
\usepackage{graphicx}	% Including figure files
\usepackage[T1]{fontenc}

\begin{document}

\title{PENTACLE: Parallelized Particle-Particle Particle-Tree Code for Planet Formation}

\author{
Masaki \textsc{Iwasawa},\altaffilmark{1,}$^{*}$
Shoichi \textsc{Oshino},\altaffilmark{2}
Michiko S. \textsc{Fujii},\altaffilmark{3}
and Yasunori \textsc{Hori}\altaffilmark{4,5}
}
\altaffiltext{1}{Advanced Institutes for Computational Science, Minatojima-minamimachi, Chuo-ku, Kobe, Hyogo 6500047, Japan}
\altaffiltext{2}{Centre for Computational Astrophysics, National Astronomical Observatory of Japan, 2-21-1 Osawa, Mitaka, Tokyo 1818588, Japan}
\altaffiltext{3}{Department of Astronomy, Graduate School of Science, The University of Tokyo, 7-3-1 Hongo, Bunkyo-ku, Tokyo 1130033, Japan}
\altaffiltext{4}{Astrobiology Centre, National Institutes of Natural Sciences, 2-21-1 Osawa, Mitaka, Tokyo 1818588, Japan }
\altaffiltext{5}{Exoplanet Detection Project, National Astronomical Observatory of Japan, 2-21-1 Osawa, Mitaka, Tokyo 1818588, Japan}
\email{masaki.iwasawa@riken.jp}


\maketitle

% Abstract of the paper
\begin{abstract}
We have newly developed a Parallelized Particle-Particle Particle-tree
code for Planet formation, {\tt PENTACLE}, which is a parallelized
hybrid $N$-body integrator executed on a CPU-based
(super)computer. {\tt PENTACLE} uses a 4th-order Hermite algorithm to
calculate gravitational interactions between particles within a cutoff
radius 
%(which is normalized by the Hill radius of a particle, $\tilde{R}_{\rm cut}$) 
and a Barnes-Hut tree method for gravity from
particles beyond. It also implements an open-source library designed
for full automatic parallelization of particle simulations, FDPS
(Framework for Developing Particle Simulator) 
to parallelize a Barnes-Hut tree algorithm for a memory-distributed supercomputer. These allow us to
handle $1-10$ million particles in a high-resolution $N$-body
simulation on CPU clusters for collisional dynamics, including
physical collisions in a planetesimal disc. In this paper, we show the
performance and the accuracy of {\tt PENTACLE} in terms of
$\tilde{R}_{\rm cut}$ and a time-step $\Delta t$. 
%(which is normalized by the orbital period divided by $2\pi$)
It turns out that the accuracy
of a hybrid $N$-body simulation is controlled through $\Delta t / \tilde{R}_{\rm cut}$
and $\Delta t / \tilde{R}_{\rm cut} \sim 0.1$ is necessary to simulate
accurately accretion process of a planet for $\geq 10^6$ years. 
For all those who interested in large-scale particle simulations,
{\tt PENTACLE} customized for planet formation will be freely available
from https://github.com/PENTACLE-Team/PENTACLE
under the MIT lisence.
\end{abstract}

%%%%%%%%%%%%%%%%%%%%%%%%%%%%%%%%%%%%%%%%%%%%%%%%%%
%%%%%%%%%%%%%%%%% BODY OF PAPER %%%%%%%%%%%%%%%%%%
\section{Introduction}

Planet formation proceeds via collisions and accumulation of planetesimals. 
Planetesimals, which are larger than neighbours,
can be predators because of stronger gravitational focusing.
During this early phase, larger planetesimals overwhelm smaller planetesimals, growing rapidly in an exponential
fashion \citep{1989Icar...77..330W}.
However, the growth of a large planetesimal, the so-called planetary embryo,
slows down because of the increase
in random motions of small planetesimals around itself by
gravitational scattering \citep{1993Icar..106..210I}. At this stage, 
planetary embryos grow oligarchically to be similar-size and then,
their orbital separations become $\sim 10$ times Hill radii as a result of orbital repulsion.
A series of
pathways toward planet formation are commonly known as runaway growth
and oligarchic growth \citep{1998Icar..131..171K}.

Collisional dynamics in a swarm of planetesimals is controlled by
their mutual gravity. Planetesimal accretion is a non-linear dynamical
process via multi-body interactions. An $N$-body simulation is an
effective means of examining dynamical behaviours of particles that
gravitationally interact each other. Gravitational force has an
infinite range, whereas gravitational interactions between particles
undergoing close encounters regulate the length of a time-step to
integrate the dynamical evolution of a collisional system.  As a
result, a direct $N$-body simulation for planet formation requires
$O(N^2)$ integrations per short time-step.

\begin{figure}
	 \begin{center}
	\includegraphics[width=80mm, bb = 0 0 726 561]{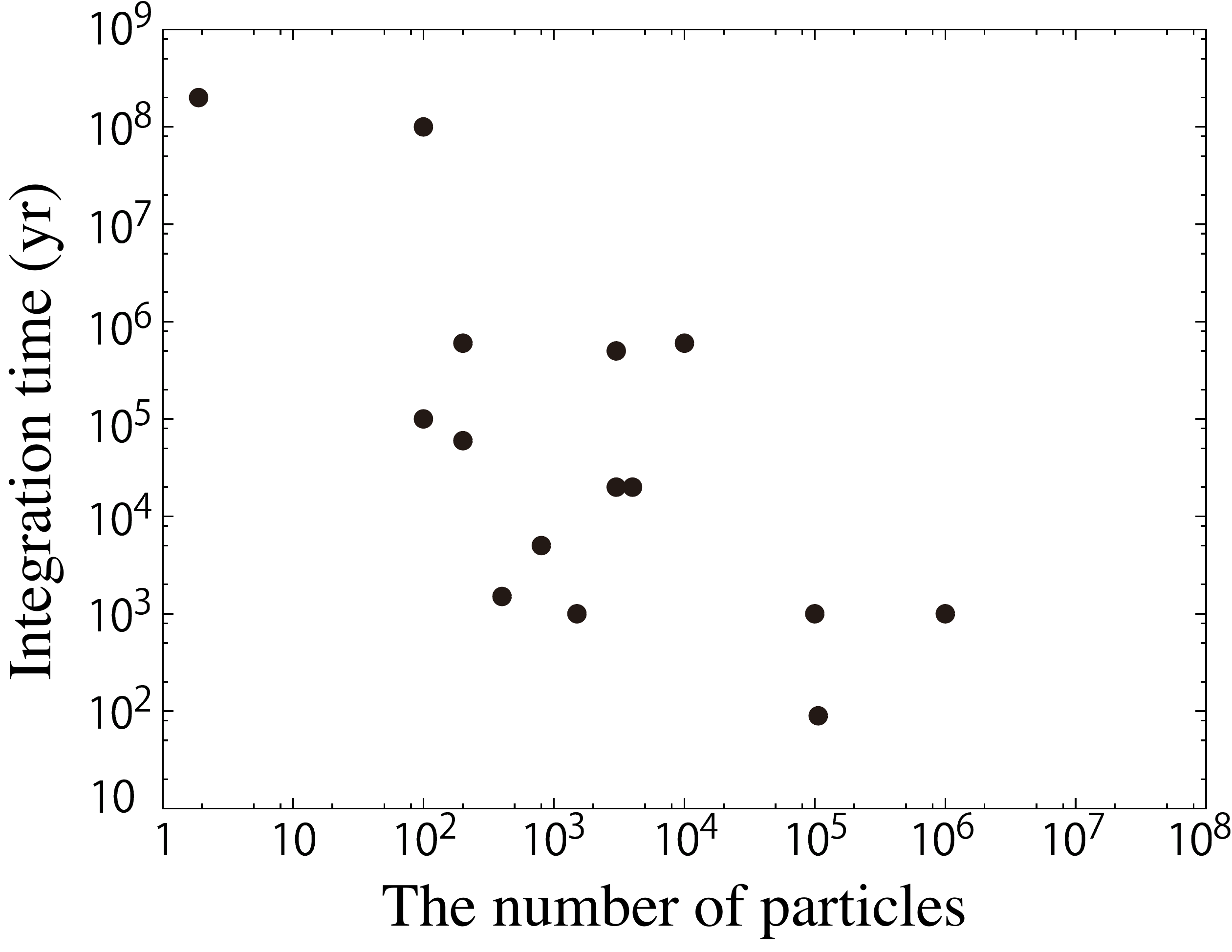}
 	\end{center}
 	\caption{History of $N$-body simulations for terrestrial
          planet formation (see \citet{2000Icar..143...45R} and
          references therein).  The current level of the number of
          particles used in $N$-body simulations of planet formation
          is $\sim 10^4-10^5$. }
	\label{fig_nbody}
\end{figure}

Due to a high computational cost of direct $N$-body simulations, we
can handle at most $\sim 10^4-10^5$ particles in simulations of
terrestrial planet formation, as shown in Fig. \ref{fig_nbody} . The
size of an equal-mass planetesimal initially assumed in such $N$-body
simulations is typically several hundred kilometres in radius, which
is similar to that of 1 Ceres and 4 Vesta. The observed population of
asteroids in the main asteroid belt suggests that the smaller
asteroids are, the more abundant they are (e.g. see Fig. 1
  in \cite{2005Icar..175..111B}). Although the present-day size
distribution of asteroids reflects the combination of a long-term
evolution of collisional processes and gravitational perturbations
from the planets, there were likely numerous small bodies at the early
stage of planet formation. In order to describe the dynamical
behaviours of small bodies in an $N$-body simulation, we need to
improve the mass resolution of particles. A classical way is to
introduce a statistical method that describes accumulation processes
of planetesimals in a sea of small bodies, following a collisional
probability; for instance, statistical simulations
(e.g. \cite{1997Icar..128..429W,2001Icar..149..235I,2011ApJ...738...35K}),
hybrid $N$-body simulations with a statistical method for small bodies
(e.g. \cite{2006AJ....131.2737B,2008Icar..198..256C}), and
super-particle approximations
\citep{2012AJ....144..119L,2015Icar..260..368M}.

Another tractable approach is to increase the number of particles
($N$) in an $N$-body simulation. There were several attempts to
accelerate and optimize processes of computing gravitational
interactions between particles.  Specialized hardwares such as HARP
and GRAPE (e.g. \cite{1990Natur.345...33S, 1993PASJ...45..349M,
  2003PASJ...55.1163M}) were developed to accelerate the calculation of
gravitational forces, dramatically increasing the number of particles
used in direct $N$-body simulations. Recently, Graphic Processing
Units (GPUs) have been introduced as an alternative accelerator
(e.g. \cite{2015MNRAS.450.4070W,2015ComAC...2....8B}). 
However, the upper limit of $N$ was still several tens of thousands due to the 
small time-steps for close encounters and relatively long 
integration time (more than one million orbital times at 1 au), which results
in a large number of steps.

In contrast,
tree methods \citep{1986Natur.324..446B} can reduce the computational
cost of gravity parts to $O(N\log N)$. 
{\tt PKDGRAV}
\citep{2001PhDT........21S} optimized for parallel computers adopts a
tree method with variable time-steps. Using {\tt PKDGRAV},
\citet{2000Icar..143...45R} simulated the dynamical evolution of a
million planetesimals for only hundreds of dynamical times.
Tree codes and also a family of particle-mesh scheme such as 
the P$^3$M scheme \citep{1981csup.book.....H}, and combinations
of the P$^3$M and tree methods 
(\cite{1995ApJS...98..355X}; \cite{2002JApA...23..185B};
 \cite{2004NewA....9..111D}; \cite{2005MNRAS.364.1105S}; 
 \cite{2005PASJ...57..849Y}; \cite{2009PASJ...61.1319I})
can treat extremely large numbers of particles, and therefore 
they are used for cosmological simulations. 
While dark matters can be considered as a collisionless system, 
planetesimal-planetesimal interactions are collisional. If cosmological $N$-body codes are applied to planetary accretion,
small time-steps for close encounters and a large number of steps become a bottleneck.

A mixed-variable symplectic (MVS) integrator, in which a Hamiltonian is
split into two parts and integrated separately, is another direction
to manipulate the calculation cost \citep{1991AJ....102.1528W,
  1991CeMDA..50...59K}.  
In a MVS integerator, gravitational interactions caused by close encounters 
are integrated using a higher-ordered scheme with smaller time 
steps, but the others are calculated using a Kepler solver or 
a fast scheme such as a tree method.
SyMBA \citep{1998AJ....116.2067D}, {\tt
  Mercury} \citep{1999MNRAS.304..793C}, and P$^3$T method
\citep{2011PASJ...63..881O} incorporate this method, albeit the way to
split a Hamiltonian is different among them.  Also, the combination of
these attempts would be possible.  {\tt GENGA}
\citep{2014ApJ...796...23G} used GPUs for computing gravity parts,
adopting a MVS method as an integrator.  \citet{2011PASJ...63..881O}
applied P$^3$T method to GRAPE and later, \citet{2015ComAC...2....6I}
developed a GPU-enabled P$^3$T method and succeeded in handling more
than a million particles, although they considered a star cluster
model in their simulation.

We newly develop a parallelized $N$-body integrator based on ${\rm
  P^3T}$ method \citep{2011PASJ...63..881O}, a Parallelized
Particle-Particle Particle-tree code for Planet formation which is
called {\tt PENTACLE} hereafter. This hybrid $N$-body code allows us
to perform a high-resolution $N$-body simulation with $1-10$ million
particles in a collisional system such as a planetesimal disc for
$\sim 1$\,Myr on a standard supercomputer.

In this paper, we present our new hybrid $N$-body code, {\tt
  PENTACLE}, in Section 2. We show the performance and accuracy of
{\tt PENTACLE} in terms of a cut-off radius and a time-step in Section
3 and also demonstrate $N$-body simulations of planetary accretion in
a swarm of 1 million planetesimals. We summarize our paper in the last
section.

\section{Methods}

In this section, we present a basic concept and algorithm of {\tt
  PENTACLE}.  In section \ref{sec:p3t}, we briefly review the
Particle-Particle Particle-Tree (${\rm P^3T}$) method which is a
hybrid symplectic integrator used in {\tt PENTACLE}.  In sections
\ref{sec:parallel_soft} and \ref{sec:parallel_hard}, we describe a
parallelization method for computing gravity forces between particles.
In the last section, we show the actual recipe of {\tt PENTACLE}.

\subsection{Particle-Particle Particle-Tree method}
\label{sec:p3t}

The concept of {\tt PENTACLE} comes from the ${\rm P^3T}$ method
\citep{2011PASJ...63..881O}.  The ${\rm P^3T}$ method is a hybrid
symplectic integrator such as SyMBA \citep{1998AJ....116.2067D} and
{\tt Mercury} \citep{1999MNRAS.304..793C}. The basic idea of this
hybrid symplectic integrator is to split a gravitational force between
two particles, i.e., a Hamiltonian, into two parts (soft and hard
parts) by their distance in heliocentric coordinates.  The Hamiltonian used in ${\rm P^3T}$ is
given by

\begin{eqnarray}
	H &=& H_{\rm hard} + H_{\rm soft}, \\
	H_{\rm hard} &=& \sum_i^N \left[ \frac{\left| {\bf p}_i \right| ^2}{2m_i} - \frac{G m_i m_0}{r_{i0}} \right]
		- \sum^N_{i} \sum_{i<j}^N\frac{G m_im_j}{{r_{ij}}}\left[ 1-W (r_{ij})\right], \\
	H_{\rm soft} &=& -\sum^N_{i} \sum_{i<j}^N\frac{G m_i m_j}{r_{ij}}W( r_{ij}), \\
	{\bf q}_{ij} &=& {\bf q}_i - {\bf q}_j, \\
	r_{ij} &=& \left|{\bf q}_{ij} \right|,
\end{eqnarray}

where $G$ is the gravitational constant, $m_i$, ${\bf p}_i$, and ${\bf
  q}_i$ are the mass, momentum and position of an $i$-th particle, and
the subscript $0$ indicates the Sun.  To connect between a short-range
and a long-range forces smoothly, we introduce a cutoff function
$W(r_{ij})$ defined as

\begin{eqnarray}
	W(y; \gamma) &=& \left\{
	\begin{array}{ll}
		\frac{7(\gamma^6 - 9\gamma^5 + 45\gamma^4 - 60\gamma^3{\rm log}\gamma - 45\gamma^2 + 9\gamma - 1)}
			{3(\gamma-1)^7}y & (y < \gamma) \\
		f(y; \gamma) + (1-f(1;\gamma))y & ( \gamma \leq y < 1) \\
		1 &  ( 1 \leq y) \\
	\end{array} \right. ,\\
	f(y;\gamma) &=& \left( -10/3y^7 + 14(\gamma+1)y^6 - 21(\gamma^2+3\gamma+1)y^5 \right. \nonumber \\
			&& + ( 35(\gamma^3+9\gamma^2+9\gamma+1)/3 )y^4 \nonumber \\
			&& - 70(\gamma^3+3\gamma^2+\gamma)y^3 \nonumber \\
			&& + 210(\gamma^3+\gamma^2)y^2 - 140\gamma^3y{\rm log}(y) \nonumber \\
			&& \left. + (\gamma^7-7\gamma^6+21\gamma^5-35\gamma^4) \right) / (\gamma-1)^7, \\
				\gamma &=& \frac{r_{\rm in}}{r_{\rm out}}, \\
			y &=& \frac{r_{ij}}{r_{\rm out}},
\end{eqnarray}

where ${r_{\rm out}}$ and ${r_{\rm in}}$ are the outer and inner
cutoff radii (see \cite{2015ComAC...2....6I}).  This cutoff function
approaches asymptotically zero if $r_{ij} < {r_{\rm in}}$ and becomes
unity if $r_{ij} > {r_{\rm out}}$.  In {\tt PENTACLE}, we use $\gamma
= 0.1$.  Forces derived from the two components of the Hamiltonian are
given by

\begin{eqnarray}
	{\bf F}_{{\rm hard},i} &=& - \frac{\partial H_{\rm hard}}{\partial {\bf q}_i} \\
        			              &=&  - \sum_{j \neq i}^N\frac{G m_im_j}{r_{ij}^3}\left[1-K(r_{ij})\right]{\bf q}_{ij}
              				           - \frac{G m_i m_0}{r_{i0}^3}{\bf q}_{i0}, \\
	{\bf F}_{{\rm soft},i} &=& - \frac{\partial H_{\rm soft}}{\partial {\bf q}_i} = - \sum_{j \neq i}^N\frac{G m_im_j}{r_{ij}^3}K(r_{ij}){\bf q}_{ij}, \\
		K(x) &=& \left\{ \begin{array}{ll} 0 &(x < 0) \\
		-20x^7 + 70x^6 - 84x^5 +35x^4 &( 0 \leq x < 1) \\
		1 &( 1 \leq x) \\
	\end{array} \right. ,\\
	x &=& \frac{y-\gamma}{1-\gamma}.
\end{eqnarray}

A formal solution of the canonical equation of motion for a given
Hamiltonian $H$ is given by

\begin{eqnarray}
	{\bf w}_i(t+\delta t) &=& e^{\delta t \{, H\}} {\bf w}_i(t) = e^{\delta t \{, H_{\rm soft} + H_{\rm hard}\}} {\bf w}_i(t),
	\label{eq_canonical} \\
	{\bf w}_i &=& \left( {\bf q}_i, {\bf p}_i \right),
\end{eqnarray}

where ${\bf w}_i$ is the canonical variable of an $i$-th particle in
phase space and $\{,\}$ means the Poisson bracket.  In {\tt PENTACLE},
we integrate time evolution of ${\bf w}_i$ with a second-order
approximation.  If a particle of which the nearest neighbour is farther
than $r_{\rm out}$, it interacts only with the Sun and its motion is
calculated by solving the Kepler equation.

For the hard part, the ${\rm P^3T}$ method adopts a fourth-order
Hermite scheme \citep{1991ApJ...369..200M} with an individual
time-step method \citep{1963MNRAS.126..223A}.  The individual
time-step method allows us to handle readily close encounters and
physical collisions. When $r_{\rm out}$ is equal to the Hill radius of
a particle, most particles have no counterpart inside $r_{\rm out}$.
The number of integration for the hard part is $O(N)$ per time-step,
whereas that for the soft part is $O(N^2)$.  Thus, we apply the
Barnes-Hut tree method \citep{1986Natur.324..446B} (with up to the
quadrupole moment) to the ${\rm P^3T}$, which reduces a computational
cost from $O(N^2)$ to $O(N{\rm log}N)$.  We also create a list of
neighbours based on a tree structure in order to evaluate efficiently
forces from the hard part.

\subsection{Parallelization of {\tt PENTACLE}}
\subsubsection{The soft part of a gravity force}
\label{sec:parallel_soft}

One of essential ingredients to perform large $N$-body simulations is
parallelization for calculating gravitational interactions between
particles.  We apply a parallelization method executed on
distributed-memory parallel computers to {\tt PENTACLE}.  This method
consists of the following steps.

\begin{enumerate}
	\item A computational domain is divided into sub-domains, each
          of which is allocated to one MPI process.
	\item Particles are assigned to each process, and a tree
          structure is constructed in it.
	\item Based on the tree structures, processes provide
          information of particles, i.e., multipole moments of a
          gravitational potential, to each other [here after this step
            is called ``exchange Local Essential Tree(LET)''].
        \item Using the information received at previous step,
          reconstruct entire tree structures on each process.
	\item Each process evaluates gravity forces between particles
          using the tree structure and integrates the motions of their
          assigned particles by using a leapfrog scheme and a fixed time-step.
\end{enumerate}

In order to efficiently implement this scheme, we use a library called
FDPS (see \cite{2016PASJ...68...54I}).  FDPS is a C++ template
library that helps users develop parallel particle-based simulation
codes.  The basic idea of FDPS is to separate technical parts involved
in parallelization from the physical problem itself: specifically, the
decomposition of a computational domain into sub-ones, exchanging
particles among inter-processes, and gathering information of
particles stored in other processes.  FDPS provides functions
necessary for parallelized tree-codes as C++ templates.  Thus, users
just define an arbitrary data structure and kernel function of a
potential between two interacting particles, and then FDPS takes care
of data exchanges among processes.  In {\tt PENTACLE}, we utilize a
cutoff function, \textbf{$\bf K(r_{ij})$}, as a kernel function of gravitational
interactions between particles.

FDPS also has APIs to search for neighbouring particles; for example,
we can use {\tt getneighbourListOneParticle()} to find particles within
the radius of $r_{\rm out}$ around a given particle.

\subsubsection{The hard part of a gravity force}
\label{sec:parallel_hard}

Parallelization of the hard part in {\tt PENTACLE} is more
straightforward.  If particles has no neighbour within $r_{\rm out}$,
their motions can be individually integrated by solving the Kepler
equation on each process. We use the Newton-Raphson method
to solve the Kepler equation in this scheme.
For convenience, we introduce the term
``cluster'' which is defined as a subset of particles. Each particle
must belongs to a cluster and clusters are exclusive each other. All
neighbour particles of an arbitrary particle in a given cluster belong
to the same cluster. In other words, a particle out of a given cluster
dose not have neighbours in this cluster. We also define the size of
cluster as the number of particles in a given cluster and refer a
cluster with the size of $k$ as $k$-cluster. Thus a particle with no
neighbours is in a 1-cluster.

When a particle has a neighbour and its neighbour is the target particle
only (i.e. the both particles are in a 2-cluster), the particle
interacts only with its neighbour and the Sun. This system can be
considered as an isolated three-body system. If both particles are
loaded on the same process, any inter-process communication is not
required. Otherwise, we just send the particle's data to the process
in which its neighbour is stored.

In principle, we can integrate the motion of a particle with multiple
neighbours in a similar way.  It, however, is a little bit complicated
to find a cluster with the size $\geq 2$ (hereafter $k_{>2}$-cluster)
in parallel. We first send particles in $k_{>2}$-cluster to the root
process named as "rank 0" and the root process integrates their
motions in serial order.  As shown later, particles in
$k_{>2}$-cluster are rare and this serial procedure has little effect
on the scaling performance of {\tt PENTACLE} for $N\lesssim 10^6$.

\subsection{Procedures in {\tt PENTACLE}}
\label{sec:procedures}

The actual calculation of an $N$-body simulation in {\tt PENTACLE}
proceeds as follows;

\begin{enumerate}
	\item Define a data structure of a particle and a kernel function of gravitational interactions between particles.
	\item Send particles' data to FDPS.
	\item FDPS returns the soft force and the number of neighbours for each particle.\label{step:get_force}
	\item Each process gives all the assigned particles their kick velocities using their soft forces.
	\item According to the number of neighbours,
		particles are classified into three groups: a non-neighbour, one-neighbour, and multiple-neighbour group.
	\item Each process integrates the motion of particles without neighbours, the so-called drift-step, by solving the Kepler equation.
	\item If a particle has a neighbour, each process checks if the neighbour's neighbour is the target particle only, i.e., a referred particle.
		If the referred and neighbour particles are assigned to the same process, 
		the two are integrated in their process, using a fourth-order Hermite scheme.
		If not, the referred particle is sent to the neighbour's process and both processes integrate their motions.
		If the neighbour has $\geq 2$ neighbours, the referred particle is sent to the process with rank 0 (the root process).
	\item Each process sends particles with $\geq 2$ neighbours to the root process.
	\item The root process integrates all the particles received.
	\item The root process returns particles' data to their original processes.
\end{enumerate}

\section{Results}

We demonstrate planetary accretion in a swarm of planetesimals by
using {\tt PENTACLE}.  Our simulations start from a system of
equal-mass planetesimals with mean density of $2\,{\rm g}\,\,{\rm
  cm}^{-3}$ in a gas-free environment.  We use the minimum mass solar
nebula model \citep{1981PThPS..70...35H} as a nominal surface density
profile of solid material, $\Sigma_{\rm solid}$, given by

\begin{equation}
  \Sigma_{\rm solid} = 10\, \eta_{\rm ice}\, \left(\frac{a}{1\,\rm{au}} \right)^{3/2}\,\,\,{\rm g}\,\,{\rm cm}^{-2},
\end{equation}
        
where $a$ is the semi-major axis and $\eta_{\rm ice}$ is the
enrichment factor of a surface density of solid material beyond a snow
line ($a_{\rm snow} = 2.7$\,au) due to ice condensation.

We consider two planetesimal disc models for benchmark tests.  One is
a narrow ring model (model R) between 0.95\,au and 1.05\,au and the
other is a disc model (model D) with radius of 1\,au -- 11\,au. The
ring width is large enough to trace motions of planetesimals spreading
out in a disc until the end of our simulations.  In order to
investigate impacts of a computational domain size on the
applicability of our new algorithm, we use $\eta_{\rm ice} = 1$.  This
leads to a total solid amount of $0.236\,M_\oplus$ for model R and
that of $10.9\,M_\oplus$ for model D (see also Table \ref{tb:models}).

An initial eccentricity and inclination of each planetesimal are given
by a Rayleigh distribution with dispersion of $\langle e^2
\rangle^{1/2} = 2\,\langle i^2 \rangle^{1/2} = 2\sqrt{2} h$, where $e$
and $i$ are the eccentricity and inclination of a planetesimal and $h$
is the reduced Hill radius defined by the ratio of the Hill radius of
a body to its semi-major axis, i.e., $h = r_{\rm H}/a$
\footnote{In the dispersion-dominated regime, a swarm of planetesimals in a Kepler potential spontaneously
		reach the isotropic state of $e = 2\,i$ through the viscous stirring
		between themselves, irrespective of the initial conditions
		(e.g. \cite{1992Icar...96..107I}).
}

\subsection{Comparison with direct $N$-body simulations}

Assuming three different random seeds for initial values of orbital
elements of planetesimals, we performed $N$-body simulations of $5
\times 10^3$ planetesimals for model R.  We adopted two different
schemes, $\tt PENTACLE$ and a forth-order Hermite scheme
\citep{1991ApJ...369..200M}.  We use $\tilde{R}_{\rm cut} = 0.3$,
$\theta = 0.5$, and $\Delta t = 1/64$, where $\tilde{R}_{\rm cut} =
r_{\rm out}/r_H$, $\theta$ is the opening angle used in the ${\rm
  P}^3{\rm T}$ method \citep{2011PASJ...63..881O},  $\Delta t$ is
the time-step whose unit is $t_{\rm kep}/2\pi$ ($t_{\rm kep}$ is the
Kepler time), and $r_H$ is the Hill radius.

\begin{figure*}
	\begin{center}
  	\includegraphics[width=160mm, bb= 0 0 730 272]{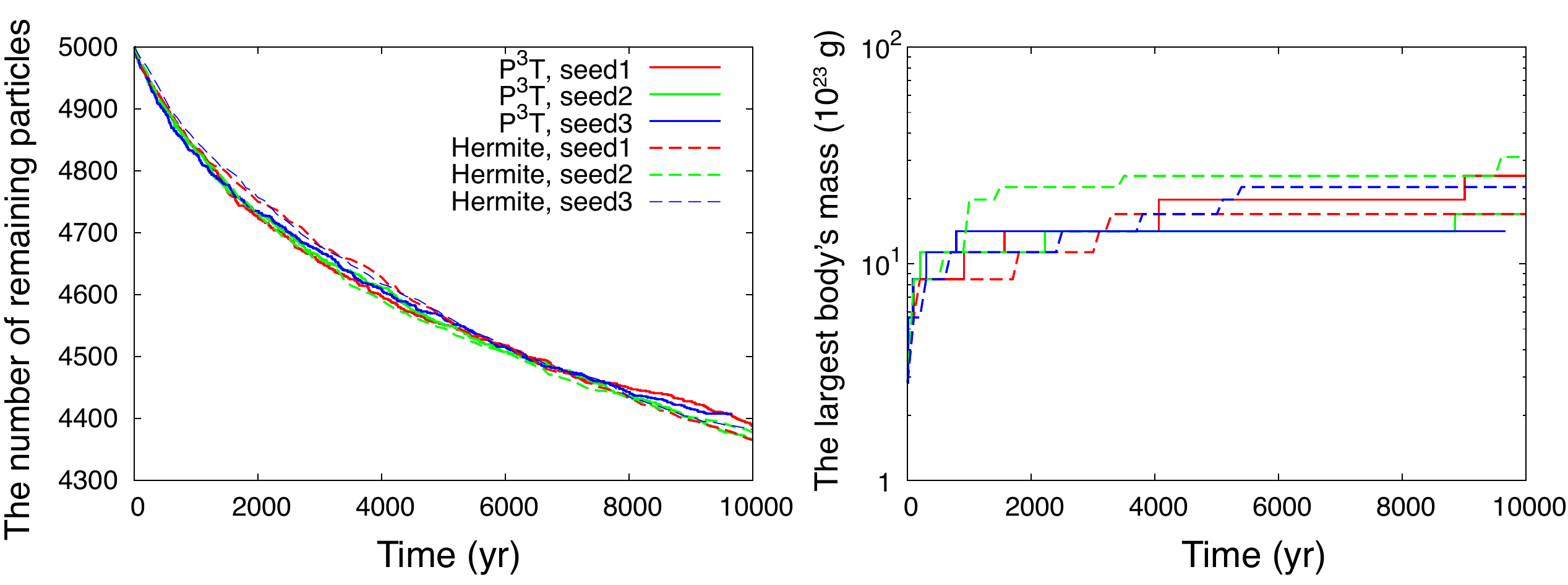}
	 \end{center}
	 \caption{$N$-body simulations of $5 \times 10^3$
           planetesimals for $10^4$\,yrs by using {\tt PENTACLE}
           (solid) and a fourth-order Hermite scheme (dashed).  Three
           colours correspond to three different random seeds for
           initial conditions of a planetesimal disc.  Left: the
           number of remaining particles. Right: mass of the largest
           body.  }
	\label{fig:comp}
\end{figure*}

\begin{figure*}
	\begin{center}
  		\includegraphics[width=160mm, bb= 0 0 690 250]{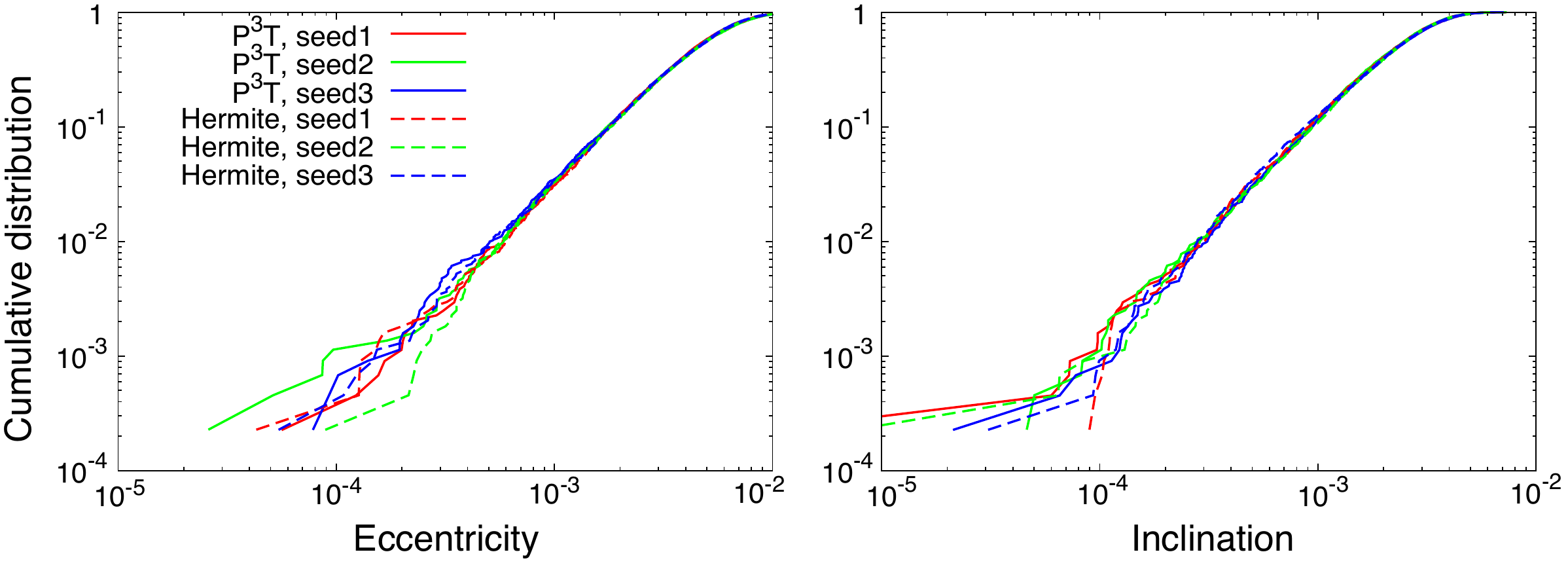}
	 \end{center}
	\caption{ Cumulative distribution functions of the eccentricity (left panel) and inclination (right panel) of planetesimals at $9\times 10^3$ year.
	        }
	\label{fig:compDF}
\end{figure*}

Figure \ref{fig:comp} shows that $\tt PENTACLE$ reproduces well results
of direct $N$-body simulations.  In all the runs, the number of
remaining planetesimals decreases monotonously in a similar manner.
On the other hand, we see car chases in the bottom panel of
Fig. \ref{fig:comp}.  Physical collisions between planetesimals are
stochastic processes.  As a result, mass evolution of the largest body
(the so-called planetary embryo) shows a stepwise growth.  In any
case, we find that a planetary embryo reaches almost the same mass
after $10^4$\,years. We also plot distribution functions of the
eccentricity and inclination of planetesimals at $9\times 10^3$ year
in Fig. \ref{fig:compDF}. There is no significant difference for all the
runs.

\subsection{Accuracy and performance}

\begin{table}
\caption{Planetesimal disc models}
%\centering
\begin{tabular}{lccc}
    \hline 
    Name &  Radius (au) & Width (au) &  Total Mass ($M_\oplus$) \\
    \hline
    Model R & 0.95 -- 1.05 &  0.1 &  0.236 \\
    Model D & 1 -- 11 &  10 &  10.9 \\
    \hline
  \end{tabular}\label{tb:models}
\end{table}

\subsubsection{Energy conservation}
\label{sec:E_conserve}

We verify energy conservation in a system of planetesimals for model R
with $N=10^6$. We consider two additional models with different
initial velocity dispersions: $\left< e^2 \right> ^{1/2} = 2 \left<
i^2 \right> ^{1/2} = 8\sqrt{2}h$ (hot disc) and $1/\sqrt{2}h$ (cold
disc).  We choose $\theta = 0.1$ and $\eta = 0.025$ in order to
suppress energy errors which arise from both a tree approximation and
a forth-order Hermite scheme, where $\eta$ is the accuracy parameter
of timesteps for the forth-order Hermite scheme
(e.g. see \cite{1992PASJ...44..141M}).

\begin{figure*}
 \begin{center}
  \includegraphics[width=120mm, bb= 0 0 719 651]{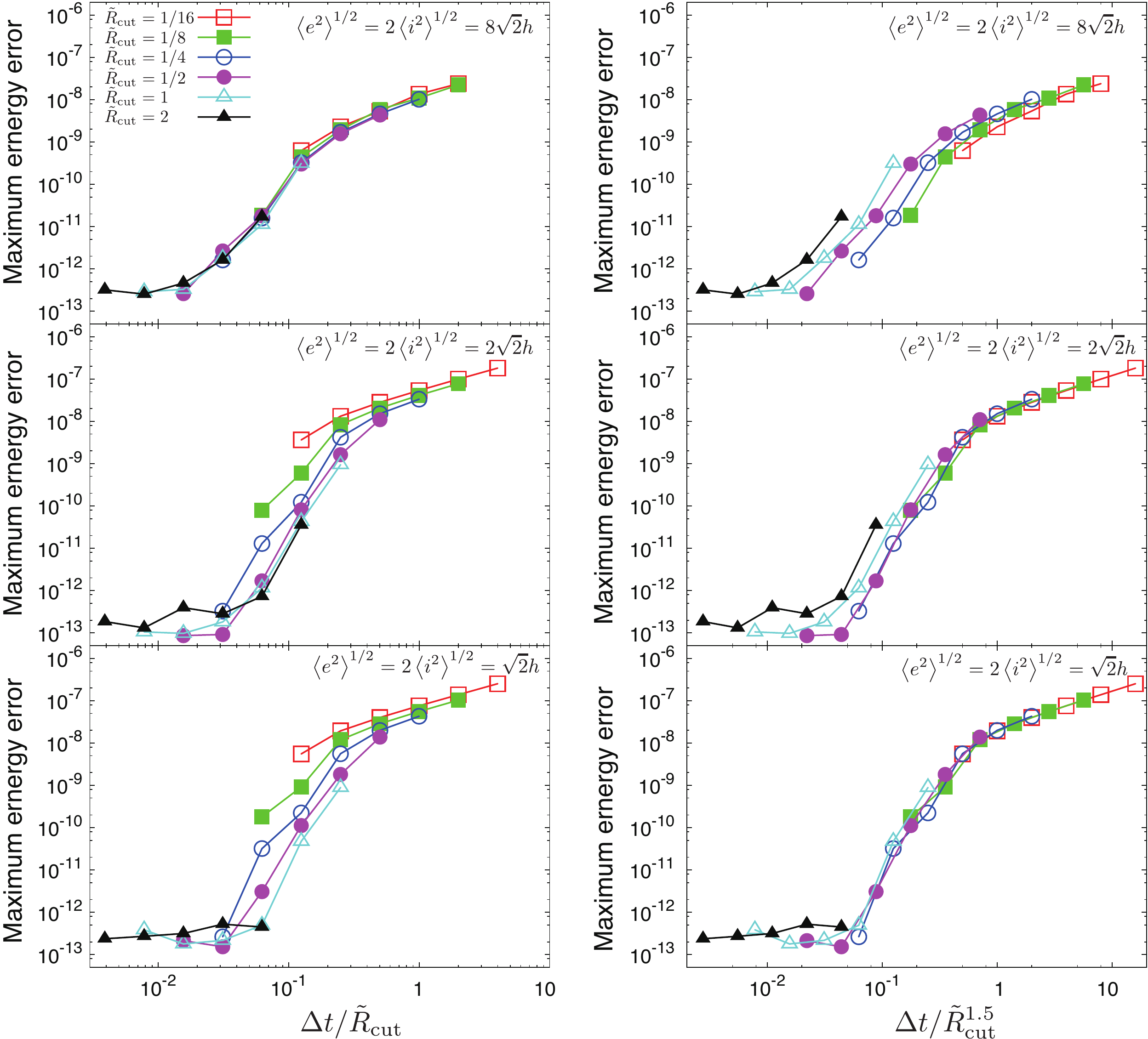}
 \end{center}
 \vspace*{0.8cm}
 \caption{ Maximum energy error over ten Keplerian orbits as functions
   of $\Delta t/\tilde{R}_{\rm cut}$ (left panel) and $\Delta
   t/(\tilde{R}_{\rm cut})^{1.5}$ (right panel). Top, middle and
   bottom panels show results of hot, standard, and cold disc
   models, respectively.}
 \label{fig:dt-Eerr}
\end{figure*}

Figure \ref{fig:dt-Eerr} shows the maximum relative energy error over 10
Keplerian orbits as functions of $\Delta t$ and $\tilde{R}_{\rm cut}$.
For the hot disc, energy errors can be controlled through $\Delta
t/\tilde{R}_{\rm cut}$.  This dependency is the same as that seen in a
system of objects with high velocity dispersions such as star cluster
simulations \citep{2015ComAC...2....6I}.  On the other hand, energy
errors rather depend on $\Delta t/(\tilde{R}_{\rm cut})^{1.5}$ than
$\Delta t/(\tilde{R}_{\rm cut})$ if the collisional system is
cold. Energy errors of a cold system mainly come from close
encounters. Since a characteristic time-scale of a close encounter is
a Keplerian time, energy errors of a cold system depends on
$\tilde{R}_{\rm cut}^{-1.5}$. However, planetesimal discs are heated
up quickly through viscous stirring and/or dynamical friction.  As a
result, we, {\it a posteriori}, can use $\Delta t/\tilde{R}_{\rm cut}$ as an
accuracy parameter for the soft part in this paper.

\begin{figure*}
 \begin{center}
  \includegraphics[width=120mm,bb= 0 0 397 214]{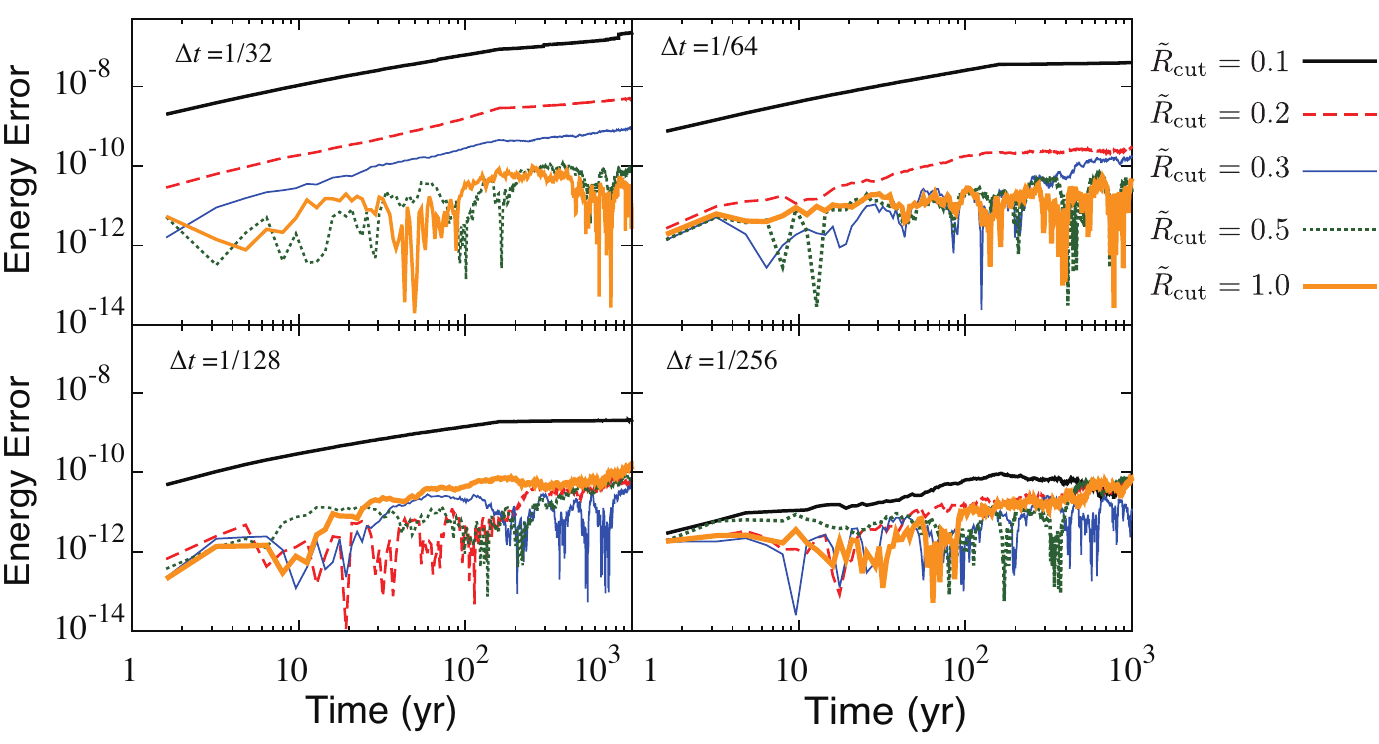}
 \end{center}
 \caption{Relative energy error as a function of time for model R
   with $N = 10^6$ and $\theta=0.5$: $\tilde{R}_{\rm cut} = 0.1$
   (solid, black), 0.2 (dashed, red), 0.3 (thin solid, blue), 0.5 (dotted, green), and 1.0 (thick solid, orange).
   From the left top to the right bottom, we use $\Delta t = 1/32,
   1/64, 1/128,\,{\rm and}\,1/256$.  }
\label{fig:energy}
\end{figure*}

Figure \ref{fig:energy} shows time evolution of relative energy errors
to 1,000 Kepler time for model R with $N = 10^6$, $\theta=0.5$, and
$\eta=0.1$ with respect to five different $\tilde{R}_{\rm cut}$. For all the runs, the
energy errors grow gradually as $t^{1/2}$, as expected for a random
walk. Thus, in order to suppress the energy error ($< 10^{-7})$ for $\sim
10^6$ years, we should chose $\Delta t/\tilde{R}_{\rm cut} \lesssim
0.1$ (see also Fig. \ref{fig:dt-Eerr}). More details about
dependencies of accuracy on parameters are discussed in
\citet{2011PASJ...63..881O}.

\begin{figure*}
 \begin{center}
  \includegraphics[width=120mm, bb= 0 0 339 232]{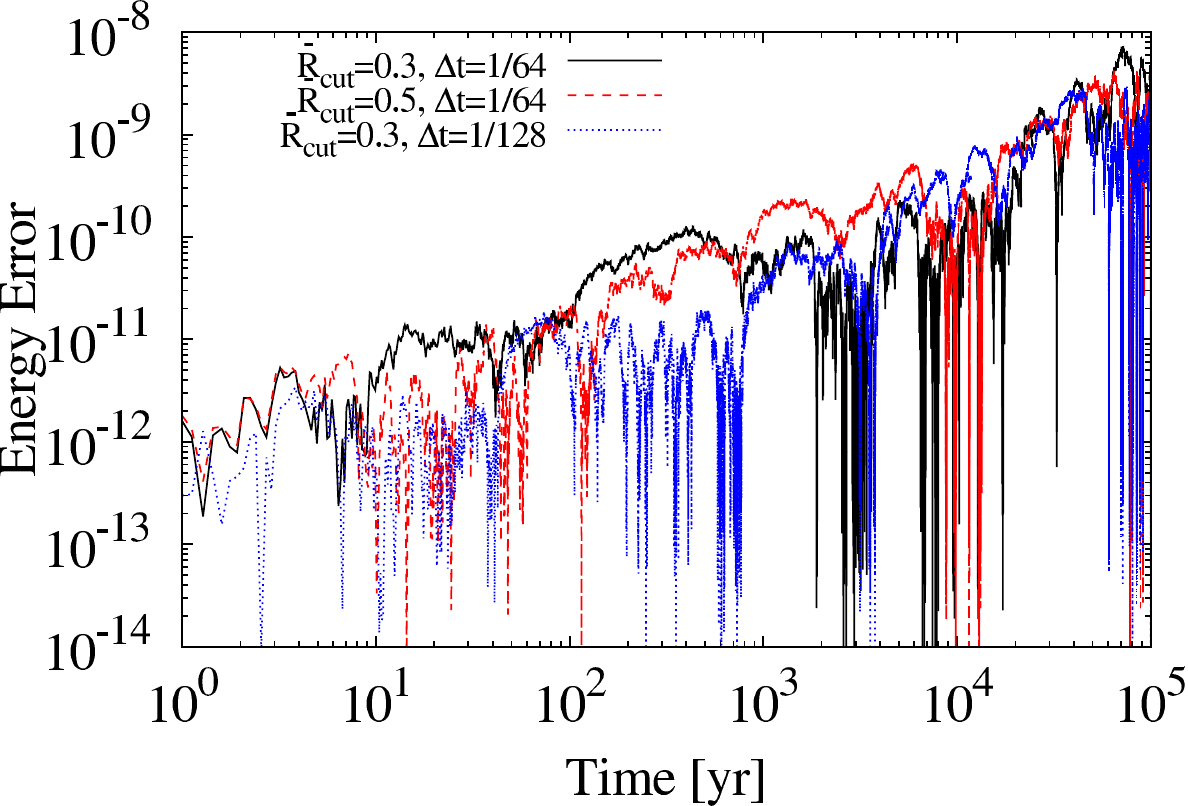}
 \end{center}
 \caption{Relative energy errors as a function of time for model R
   with $N = 10^6$ and $\theta=0.5$: $\tilde{R}_{\rm cut} = 0.3, \Delta t=1/64$
   (solid, black), $\tilde{R}_{\rm cut} = 0.5, \Delta t=1/64$ (dashed, red), and 
   $\tilde{R}_{\rm cut} = 0.3, \Delta t=1/128$ (dotted, blue).  }
\label{fig:long}
\end{figure*}

Figure \ref{fig:long} shows time evolution of relative energy errors
to 100,000 Kepler times for model R with $N = 10^6$, $\theta=0.5$, and
$\eta=0.1$. We chose three different parameter sets of $\tilde{R}_{\rm cut}$ and $\Delta t$,
such as $\tilde{R}_{\rm cut} = 0.3, \Delta t=1/64$, $\tilde{R}_{\rm cut} = 0.5, \Delta t=1/64$, 
and   $\tilde{R}_{\rm cut} = 0.3, \Delta t=1/128$. 
These results show that the increase in energy errors roughly obeys $t^{1/2}$ over a long period of time.

\subsubsection{Effect of disc radius}

\begin{figure*}
 \begin{center}
  \includegraphics[width=80mm, bb=0 0 348 250]{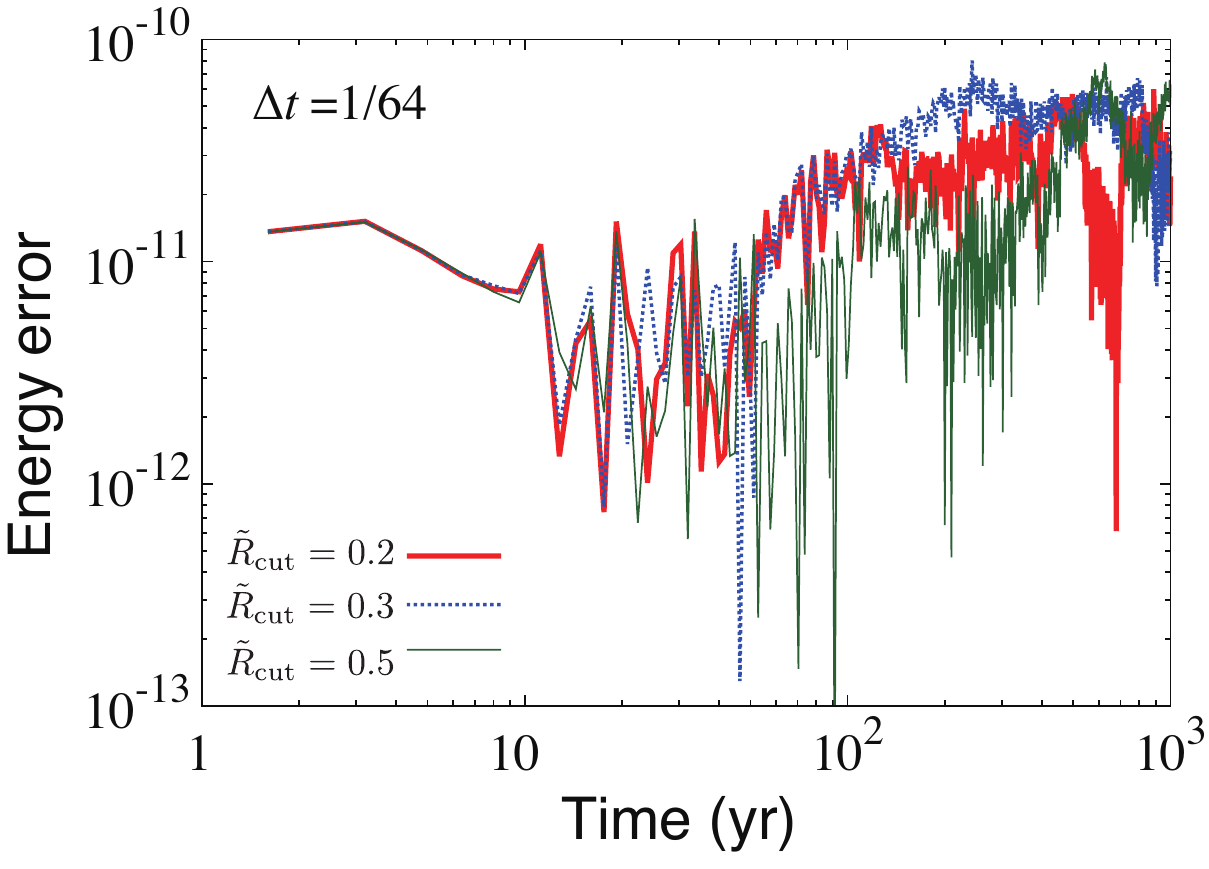}
  \includegraphics[width=80mm,bb=0 0 350 252]{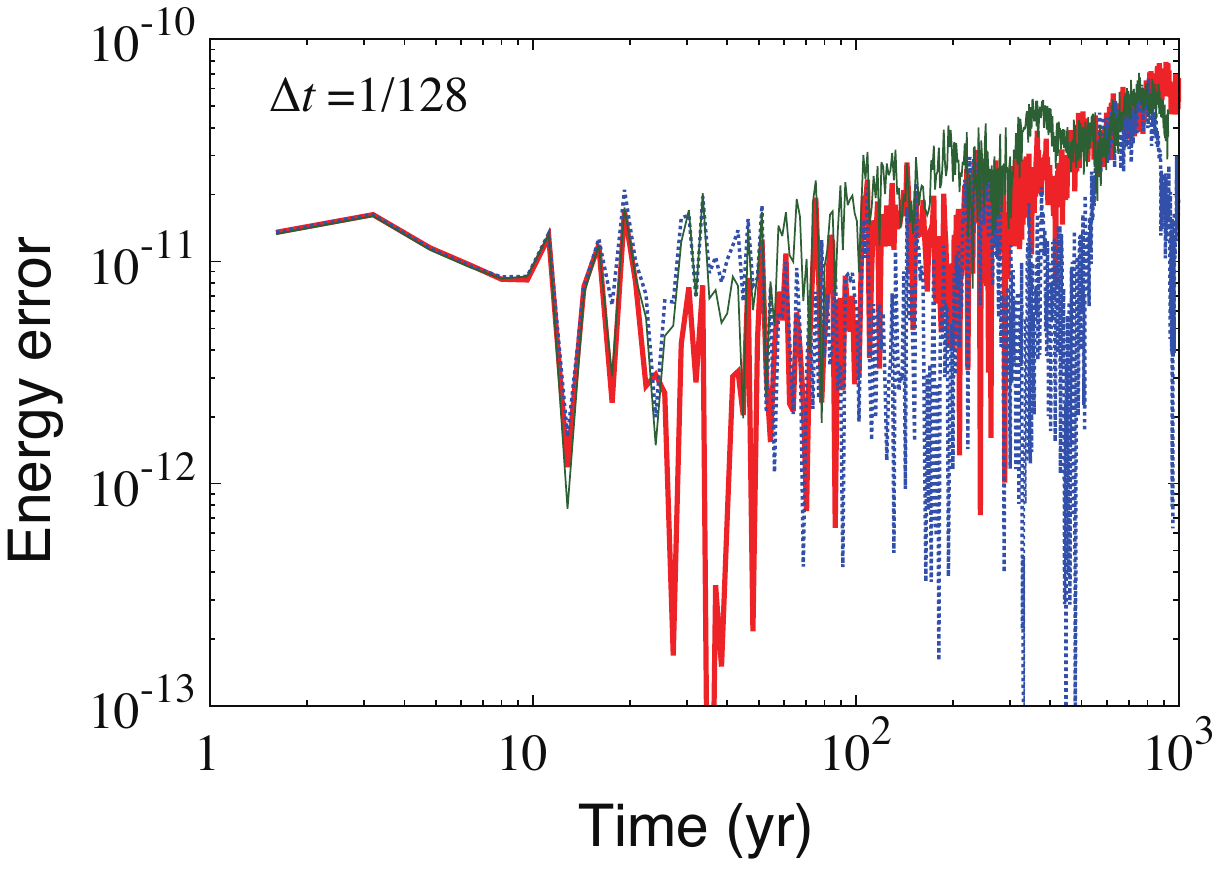}
 \end{center}
 \caption{Energy errors for model D with $N =10^6$
 		for $\Delta t=1/64$ (left) and $1/128$ (right): $\tilde{R}_{\rm cut} = 0.2$
   (thick solid, red), 0.3 (dotted, blue), and 0.5 (thin, green).
		}
		\label{fig:disc}
\end{figure*}

In order to investigate an effect of a disc radius on energy
conservation, we also perform simulations similar to those in
section \ref{sec:E_conserve} for model D with $N = 10^6$, $\theta =
0.5$, and $\tilde{R}_{\rm cut} = 0.2, 0.3,\,{\rm and}\,0.5$. We adopt
$\Delta t=1/64$ and $1/128$ for these runs. The energy error is 
shown in Fig. \ref{fig:energy}, and the behaviour is similar to 
that of model R (see Fig. \ref{fig:disc}). This means that 
{\tt PENTACLE} is applicable to $N$-body
simulations for planet formation in a variety of disc models.

\subsubsection{Efficiency for parallelization}

We also discuss the performance and parallelization efficiency of
{\tt PENTACLE}. All the simulations in this section are carried out on Cray XC30
(ATERUI) at the Centre for Computational Astrophysics of the National
Astronomical Observatory of Japan. ATERUI consists of 1,060 computer
nodes and each node has two Intel Xeon E5-2690 v3 (12 cores, 2.6GHz)
processors. We assigned one CPU core to one MPI process (i.e. the
number of CPU cores $N_c$ means that of MPI processes).

\begin{figure*}
	\begin{center}
  	  \includegraphics[width=120mm,bb=0 0 1025 264]{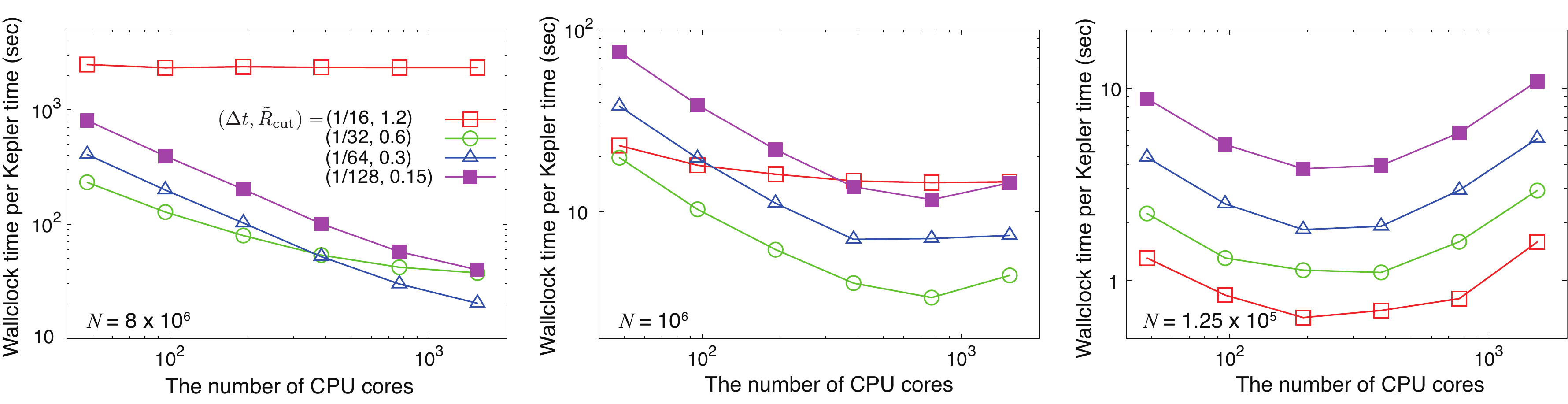}
	\end{center}
	\caption{Execution time per Kepler time for $N=8\times 10^6$
          (left panel), $10^6$ (middle panel), and $1.25\times 10^5$
          (right panel) against the number of CPU cores. Symbols
          correspond to different $\Delta t$ and $\tilde{R}_{\rm
            cut}$.}
	\label{fig:strong_scaling}
\end{figure*}

\begin{figure*}
	\begin{center}
  		\includegraphics[width=120mm,bb=0 0 1026 266]{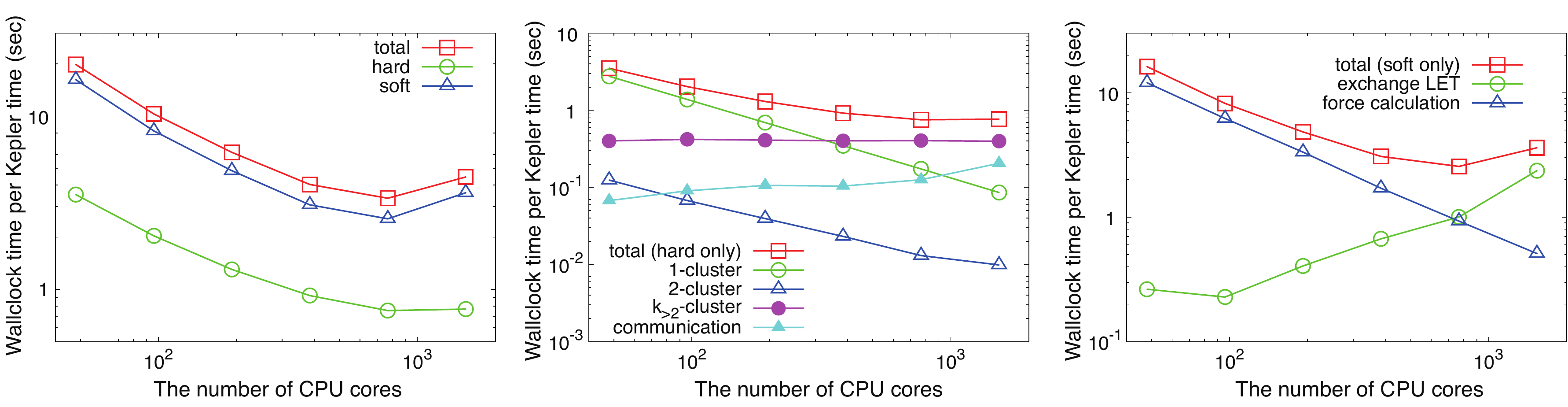}
	 \end{center}
	\caption{Breakdown of the execution time for the runs with
          $N=1$M, $\tilde{R}_{\rm cut}=0.6$ and $\Delta t=1/32$. Left
          panel: breakdown of the total execution time against the
          number of CPU cores. Open squares, open circles, and open
          triangles indicate the total execution time, the hard-part
          one, and the soft-part one, respectively.  Middle panel:
          breakdown of the execution time for the hard-part. Open
          squares indicate the total execution time for the hard part
          only. Open circles, open triangles and filled circles
          indicate the time for integrating particles in 1-, 2- and
          $k_{>2}$-cluster, respectively.  Filled triangles indicate
          the time for both gathering and scattering particles in
          $k_{>2}$-cluster from/to the root processes. Right panel:
          breakdown of the execution time for the soft-part. Open
          squares indicate the total execution time for the soft part only and
          open circles and open triangles indicate the time for the
          exchange LET and the force calculation using the tree method,
          respectively.}
	\label{fig:break_down}
\end{figure*}

Fig. \ref{fig:strong_scaling} shows the execution time per Kepler time
for various $N$ ($8\times 10^6, 10^6,$ and $1.25\times 10^5$). 
Each curve indicates different parameter sets of
$\Delta t$ and $\tilde{R}_{\rm cut}$. For all the runs, we take $\Delta t
/ \tilde{R}_{\rm cut}$ to be constant so that the energy errors of different
runs are almost the same. We see similar trends among most runs: the execution time
decreases linearly for small $N_c$ and increases for large $N_c$. To
see more details, we plot the breakdown of the execution time for the
run with $N=10^6$ (1M), $\tilde{R}_{\rm cut}=0.6$, and $\Delta t=1/32$ in
Fig. \ref{fig:break_down}. We see that the execution time for both the hard
and soft parts decrease for small $N_c$. However, as $N_c$ increase,
slopes for both parts become shallower and finally, the execution time
for the hard (soft) part stalls (increases).
For the hard part, we find a good scaling of the execution time
for integrating particles in 1- and
2-clusters, whereas that for integrating particles in
$k_{>2}$-clusters is independent of $N_c$. This is because the integration 
of $k_{>2}$-clusters is not executed in parallel. The time for 
gathering and scattering particles in $k_{>2}$-clusters from/to the 
root process is almost constant, because {\tt MPI\_Gatherv} and 
{\tt MPI\_Scatterv} used for gathering and scattering of particles are limited to the injection
bandwidth. Thus, the minimum time for the hard part is limited to
integrating or gathering (and scattering) particles in
$k_{>2}$-cluster.
For the soft part, the execution time for calculating a gravity force decreases
linearly for all the range of $N_c$. However, the time for the exchange
LET increases for large $N_c$. This is because {\tt MPI\_Alltoallv} is
used in FDPS at this step \citep{2016PASJ...68...54I}.

\begin{figure*}
	\begin{center}
  		\includegraphics[width=120mm,bb=0 0 1025 266]{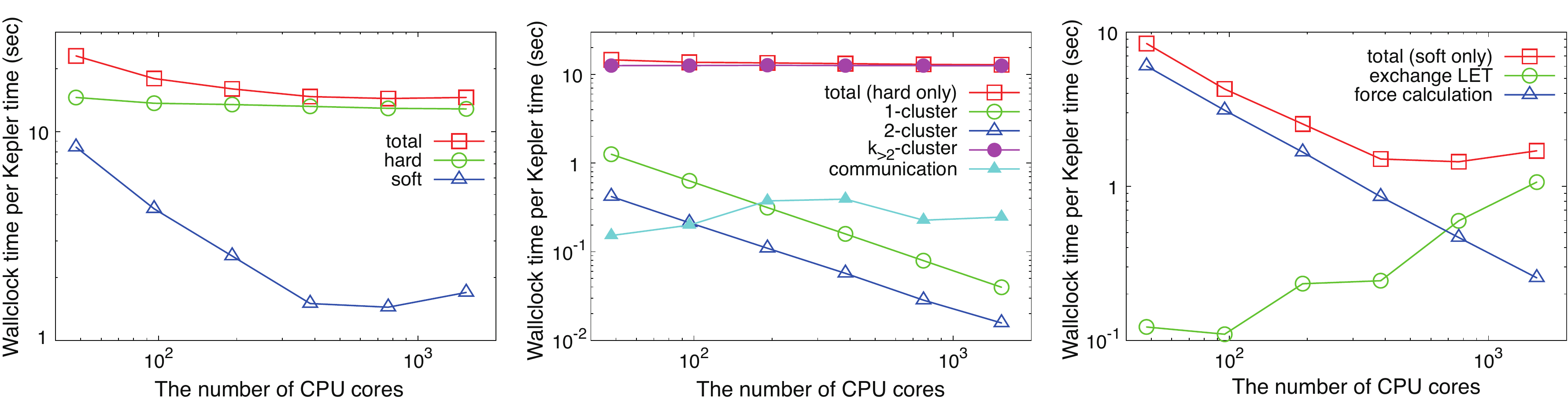}
	 \end{center}
	\caption{The same figures as Fig. \ref{fig:break_down}, but with
          $\tilde{R}_{\rm cut}=1.2$ and $\Delta t=1/16$.}
	\label{fig:break_down2}
\end{figure*}

\begin{figure*}
	\begin{center}
  		\includegraphics[width=120mm,bb=0 0 1046 266]{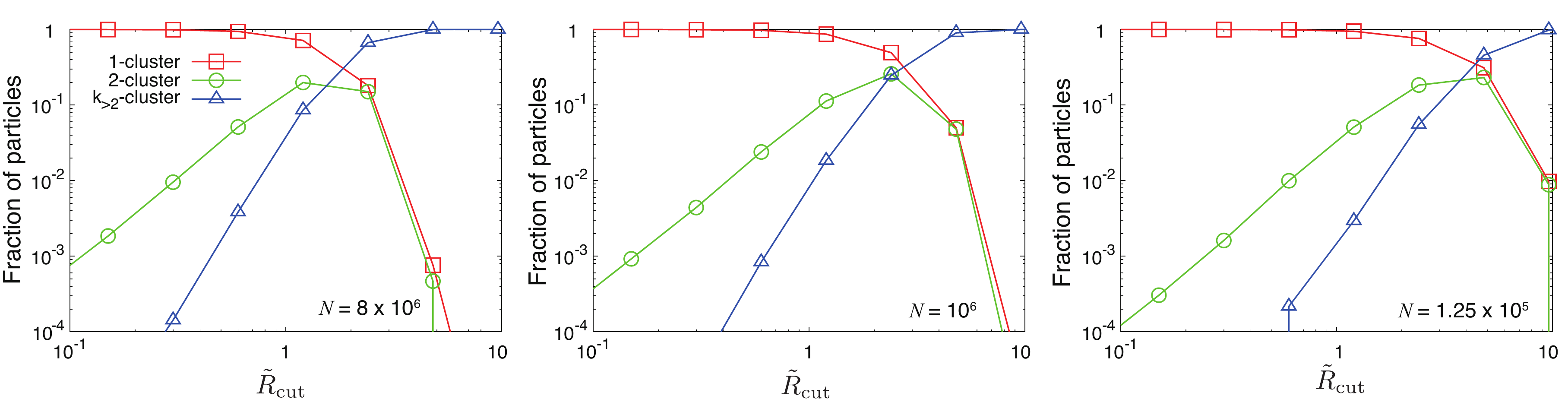}
	 \end{center}
	\caption{Fraction of particles in 1- (red open square), 2-
          (green open circle) and $k$-clusters (blue open triangle)
          against $\tilde{R}_{\rm cut}$ for the runs with $N=8\times
          10^6$ (left panel), $10^6$ (middle panel), and $1.25\times
          10^5$ (right panel).}
	\label{fig:r-np}
\end{figure*}

We also show the same figure as Fig. \ref{fig:break_down}, but with
$\tilde{R}_{\rm cut}=1.2$ and $\Delta t=1/16$. behaviours of all the
curves are similar those shown in Fig.
\ref{fig:break_down2}, except that the hard
part dominates the total execution time.
Compared with the run with $\tilde{R}_{\rm cut}=0.6$ and $\Delta
t=1/32$, more particles belong to $k_{>2}$-clusters in this case,
in other words, many particles are not integrated in parallel.
Figure \ref{fig:r-np} shows that the fraction of particles in 1-, 2-, 
and $k_{>2}$-clusters, and it indicates that the number of particle in
$k_{>2}$-cluster is about one hundred times larger than that in
$1$-cluster. If, in the hard part, each execution time of integration
of a particle per unit time is the same among all particles, the total
execution time for the hard part can not be scaled for $N_c \gtrsim
100$. This conclusion is firmly supported by Fig. \ref{fig:break_down2}.

The  execution time for the soft part of the run with $\tilde{R}_{\rm cut}=1.2$
and $\Delta t=1/16$ is roughly half of that with $\tilde{R}_{\rm
  cut}=0.6$ and $\Delta t=1/32$. If we fully parallelize
the hard part, its execution time is much
smaller than the soft one. As a result, the total execution time of the run with
$\tilde{R}_{\rm cut}=1.2$ and $\Delta t=1/16$ roughly becomes half of
that with $\tilde{R}_{\rm cut}=0.6$ and $\Delta t=1/32$.

Finally, we discuss a possibility of full parallelization of the hard part
to permit a larger $\tilde{R}_{\rm cut}$ and
$\Delta t$, namely, we reduce the number of the use of group
communication per unit time. To discuss the possibility of full
parallelization in the hard part, we plot the number of particles
in the largest cluster against the mean number of neighbours per
particle, $\left< n_{\rm ngb} \right>$ in Fig. \ref{fig:np-nc} and the
distribution function of a cluster size, i.e., $k$-cluster, in Fig.
\ref{fig:k-nc}. As seen in Fig. \ref{fig:np-nc}, there is a threshold at
$\left< n_{\rm ngb} \right> \sim 1$.

Under the circumstance of $\left< n_{\rm ngb} \right>
\lesssim 1$, many clusters contain a few
particles (see also Fig. \ref{fig:k-nc}).
This means that we could integrate the particles in $k_{\rm > 2}$-cluster in parallel
if we chose small enough $\tilde{R}_{\rm cut}$ to guarantee $\left< n_{\rm ngb} \right> < 1$.
In addition, if we could integrate all the particles for the hard part in
parallel, gathering (scattering) particles in $k_{>2}$-cluster to
(from) the root process were not necessary \footnote{Inter-adjacent node
  communications are still required but these costs would be much
  smaller than the collective one}.
Figure \ref{fig:k-nc} also shows that the distribution function of a $k$-cluster depends on $\left< n_{\rm ngb}
\right>$ and is independent of $N$.
The larger $N$ we use, the more important the parallelization of
the hard part becomes. Thus, we
will modify {\tt PENTACLE} to handle the hard part in fully parallel
in the near future.

\section{Terrestrial planet formation}

In the previous section, we demonstrated that a high-resolution $N$-body simulation with
one million particles for planet formation is doable if we use {\tt PENTACLE} code.
How does the growth of a planetary embryo proceed in a sea of "small" planetesimals?
In order to investigate effects of the initial size of a planetesimal $r_{\rm pls}$,
we carried out three $N$-body simulations of terrestrial planet formation using $N = 10^4$, $10^5$, and $10^6$,
which correspond to $r_{\rm pls} \sim 507\,$km, $235$\,km, and $109$\,km, respectively.
We considered the same narrow planetesimal ring model as that used in Section 3.1.
In these simulations, we introduce an aerodynamic drag force as described in \cite{1976PThPh..56.1756A}.
We assumed that an initial gas density at 1\,au, $\rho_0 = 2.0 \times 10^{-9}\, \mathrm{g\, cm^{-3}}$.
The density of a disc gas decreases exponentially with time in a manner of $\rho_{\rm gas} =\rho_0 \exp^{-t/{1\,\rm Myrs}}$,
where $\rho_{\rm gas}, t$ is the density of a disc gas and the time, respectively.
We simulated planetesimal accretion onto a growing planetary embryo for 1\,Myrs.

\begin{figure*}
	\begin{center}
  		\includegraphics[width=80mm,bb=0 0 341 253]{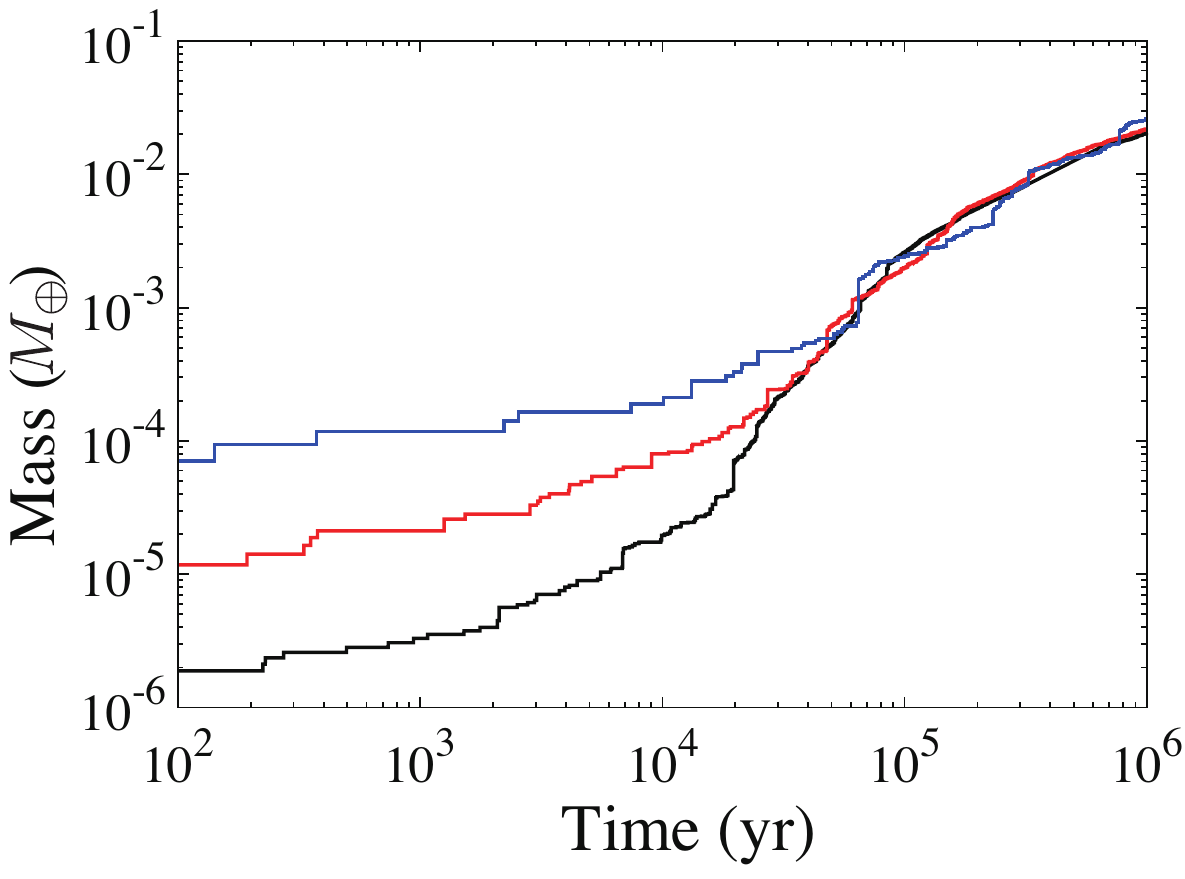}
	 \end{center}
	 \caption{Mass evolution of the largest planetary embryo in $N$-body simulations with $N = 10^4$(blue), $10^5$(red), and $10^6$(black) planetesimals, which correspond to $r_{\rm pls} \sim 507\,$km, $235$\,km, and $109$\,km, respectively.}
	 \label{fig:runaway_growth}
\end{figure*}

\begin{figure*}
	\begin{center}
  		\includegraphics[width=80mm,bb=0 0 335 259]{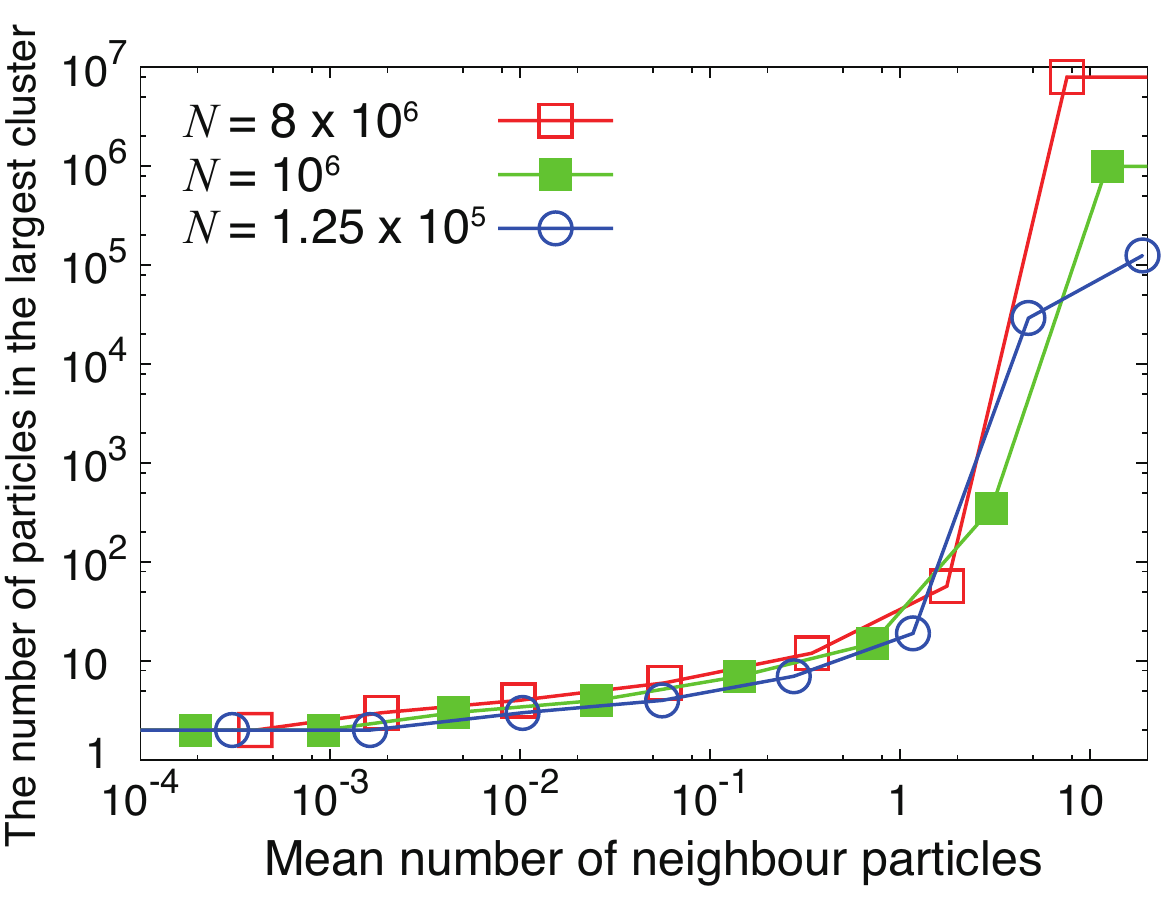}
	 \end{center}
	\caption{The number of particles in the largest cluster
          against the mean number of neighbours per particle for the
          runs with $N=8\times 10^6$ (red open square), $10^6$ (green
          filled square), and $1.25\times 10^5$ (blue open circle).}
	\label{fig:np-nc}
\end{figure*}

\begin{figure*}
	\begin{center}
  		\includegraphics[width=80mm,bb=0 0 343 254]{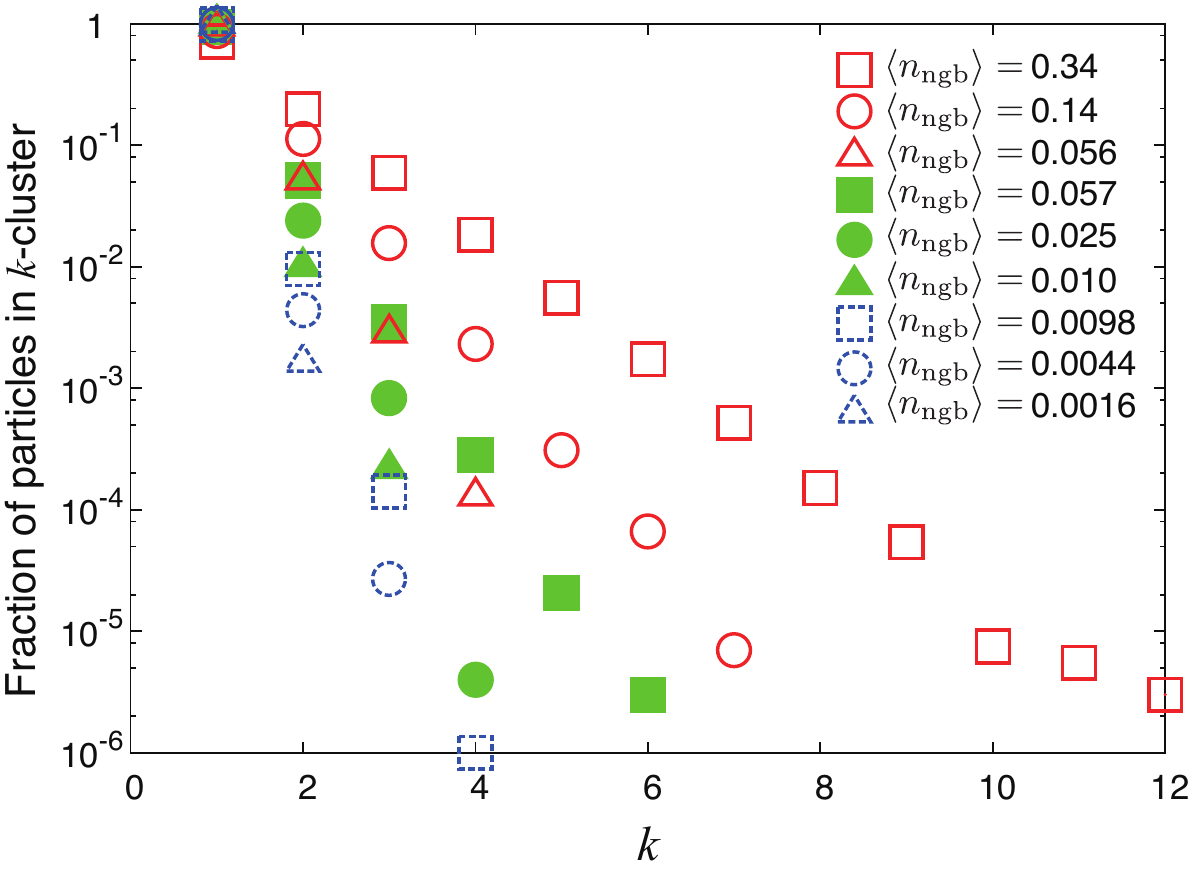}
	 \end{center}
	\caption{Fraction of particles in a $k$-cluster for runs with
          $\tilde{R}_{\rm cut}=1.2$ (red, open), $0.6$ (green, full), and $0.3$ (blue, dashed)
          for $N = 8\times 10^6$ (square), $10^6$ (circle), and $1.25\times 10^5$ (triangle).}
	\label{fig:k-nc}
\end{figure*}

Figure \ref{fig:runaway_growth} shows mass evolution of a largest body for 1\,Myrs.
We can see that a planetary embryo grows rapidly in a runaway fashion and its
growth rate follows a power-law function of mass, as shown in \citet{1993Icar..106..210I}. 
Then, the emergent runaway body enters the so-called oligarchic growth mode and 
eventually, the final mass of the embryo approaches the almost same value in the three cases.
We confirmed that a classical picture of terrestrial planet formation holds true for a swarm of 
equal-sized planetesimals with radius of $\sim 100$\,km.
A significant difference among the three cases is, however, the duration of a transient phase between the
runaway growth and the oligarchic growth. 
In the transient phase, the growth of a runaway body slows down but random velocities of ambient planetesimals are not excited so efficiently by itself yet. 
This may indicates that a planetary growth in a sea of smaller planetesimals proceeds in a non-equilibrium oligarchic growth mode, as suggested by \citet{2010ApJ...714L.103O} and \citet{2010Icar..209..836K}.
This topic will be discussed in our forthcoming paper.
Last but not the least, although energy errors of the high-resolution $N$-body simulation are not shown in this paper, they
are always lower than $10^{-7}$. 

\section{Discussions and Summary}

We have developed a parallelized hybrid $N$-body code ({\tt PENTACLE})
to perform high-resolution simulations of planet formation.  
The open-source library designed for full automatic parallelization
of particle simulations (FDPS) is implemented in {\tt PENTACLE}.  {\tt PENTACLE} uses a 4th-order Hermite scheme to calculate gravitational
interactions from particles within a cutoff radius ($\tilde{R}_{\rm cut}$) and the Barnes-Hut tree method for gravity from particles
beyond. We also confirmed that results of planetary growth in a planetesimal ring using {\tt PENTACLE} are in good agreement with
those using a direct $N$-body code.

We figured out that $\Delta t / \tilde{R}_{\rm cut} \lesssim
0.1$ reduces energy errors to an acceptable level when simulating
planetary accretion in a swarm of planetesimals (see Figs.
\ref{fig:dt-Eerr} -- \ref{fig:disc}). {\tt PENTACLE} allows us to
handle 1 -- 10 million particles in a collisional system; for example,
on a supercomputer with $10^3$ CPU cores, it takes one month to trace
the dynamical evolution of $10^6$ planetesimals for 1\,Myr.
Nevertheless, as shown in Fig. \ref{fig:strong_scaling}, the
computational cost of runs with $10^7$ or more particles would be
still expensive. The bottleneck of the performance for larger $N_{\rm c}$ (the number of CPU cores) is the
exchange LET used for {\tt MPI\_Alltoallv}. In order to relax
this problem, we could reduce the number of processes due to the usage of accelerators such as GRAPE, GPU or
PEZY-SC, while keeping the peak performance. We are now developing a GPU-enable version, {\tt
  PENTAGLE} (a Parallelized Particle-Particle Particle-tree code for
Planet formation on a GPU cluster) based on
\citet{2015ComAC...2....6I}.

We have focused the scope of this paper on the performance and
accuracy of {\tt PENTACLE}, but have also demonstrated the mass
evolution of a planetary embryo embedded in a disc with $10^6$
planetesimals.  Increasing the number of particles used in an $N$-body
simulation decreases their sizes, namely the physical size of a
planetesimal.  A random velocity distribution of planetesimals in both
a runaway and oligarchic growth mode could become different.  This
means that the growth of a planetary embryo in a sea of small
planetesimals might proceed unlike a traditional picture of
terrestrial planet formation.  This topic will be discussed in our
next paper.

{\tt PENTACLE} will be freely available under the MIT lisence for all
those who interested in large-scale particle simulations.
The source code is hosted on the GitHub platform and can be
downloaded from https://github.com/PENTACLE-Team/PENTACLE.
The next version will be exexcutable on a GPU cluster and also we will include effects of disc-planet interactions and a statistical treatment of
collisional fragmentation into {\tt PENTACLE}.

\bigskip

\begin{ack}
This project was supported by JSPS KAKENHI Grant Numbers 15H03719.
Y.H. and M.F. were supported by JSPS KAKENHI Grant Numbers 26800108
and 26103711, respectively. $N$-body simulations in this paper were
carried out on Cray XC30 (ATERUI) at the Centre for Computational
Astrophysics of the National Astronomical Observatory of Japan. This
research used computational resources of the K computer provided by
the RIKEN Advanced Institute for Computational Science through the
HPCI System Research project (Project ID:ra000008). Part of the
research covered in this paper research was funded by MEXT's program
for the Development and Improvement for the Next Generation Ultra
High-Speed Computer System, under its Subsidies for Operating the
Specific Advanced Large Research Facilities.
\end{ack}

\end{document}